\documentclass[10pt,twocolumn,letterpaper]{article}

\usepackage{iccv}
\usepackage{times}
\usepackage{epsfig}
\usepackage{graphicx}
\usepackage{amsmath, eucal}
\usepackage{amssymb, url, makecell}
\usepackage{multirow, url, color}

\usepackage[margin=0pt,font={it,footnotesize},labelfont={rm,sc},tableposition=top]{caption}

\usepackage[pagebackref=true,breaklinks=true,letterpaper=true,colorlinks,bookmarks=false]{hyperref}

\iccvfinalcopy 

\def\iccvPaperID{2975} 

\ificcvfinal\fi
\begin{document}

\title{Semi-Global Weighted Least Squares in Image Filtering}

\author{Wei Liu$^{1}$\thanks{Part of this work was  done when W. Liu was visiting The
    University of Adelaide. 
    } ~ ~ ~
    Xiaogang Chen$^2$ ~ ~ ~  Chuanhua Shen$^{3}$ ~ ~ ~  Zhi Liu$^{4}$ ~ ~ ~ Jie Yang$^{1}$\\
$^1$Shanghai Jiao Tong University,  China
~ ~ ~
$^2$University of Shanghai for Science and Technology,  China \\
$^3$The University of Adelaide, Australia ~ ~ ~ ~ ~ ~
$^4$Shanghai University, China
}

\maketitle
\thispagestyle{empty}

\begin{abstract}

Solving the global method of Weighted Least Squares (WLS) model in image filtering is both time- and memory-consuming. In this paper, we present an alternative approximation in a time- and memory- efficient manner which is denoted as Semi-Global Weighed Least Squares (SG-WLS). Instead of solving a large linear system, we propose to iteratively solve a sequence of subsystems which are one-dimensional WLS models. Although each subsystem is one-dimensional, it can take two-dimensional neighborhood information into account due to the proposed special neighborhood construction. We show such a desirable property makes our SG-WLS achieve close performance to the original two-dimensional WLS model but with much less time and memory cost. While previous related methods mainly focus on the 4-connected/8-connected neighborhood system, our SG-WLS can handle a more general and larger neighborhood system thanks to the proposed fast solution. We show such a generalization can achieve better performance than the 4-connected/8-connected neighborhood system in some applications. Our SG-WLS is $\sim20$ times faster than the WLS model. For an image of $M\times N$, the memory cost of SG-WLS is at most at the magnitude of $\max\{\frac{1}{M}, \frac{1}{N}\}$ of that of the WLS model. We show the effectiveness and efficiency of our SG-WLS in a range of applications. The code is publicly available at: \footnotesize{\color{red}\url{https://github.com/wliusjtu/Semi-Global-Weighted-Least-Squares-in-Image-Filtering}}.

\end{abstract}

\vspace{-2em}
\section{Introduction}
\label{SecIntroduction}

Image smoothing is an important operation in both image processing and computer graphics. Many applications require decomposing an image into a piecewise smooth base layer which contains the main structure information and a detail layer which captures the residual details. To achieve the decomposition, Edge-Preserving Smoothing (EPS) is required.
EPS can be achieved with \emph{local filters} which compute the output as a weighted average of the input. Bilateral filter \cite{tomasi1998bilateral} is one of the well-known filters which has been widely used in various applications such as image upsampling \cite{kopf2007joint}, flash/no flash image filtering \cite{petschnigg2004digital} and HDR tone mapping \cite{durand2002fast}. There are also other local filters based on different theories and computational models \cite{gastal2011domain, he2013guided, lu2012cross, yang2012recursive}. Most local filters can be efficiently computed. However, they can cause gradient reversals and halo artifacts \cite{farbman2008edge, he2013guided} which are their main drawbacks.

\begin{figure}
\centering
  \includegraphics[width=0.9\linewidth]{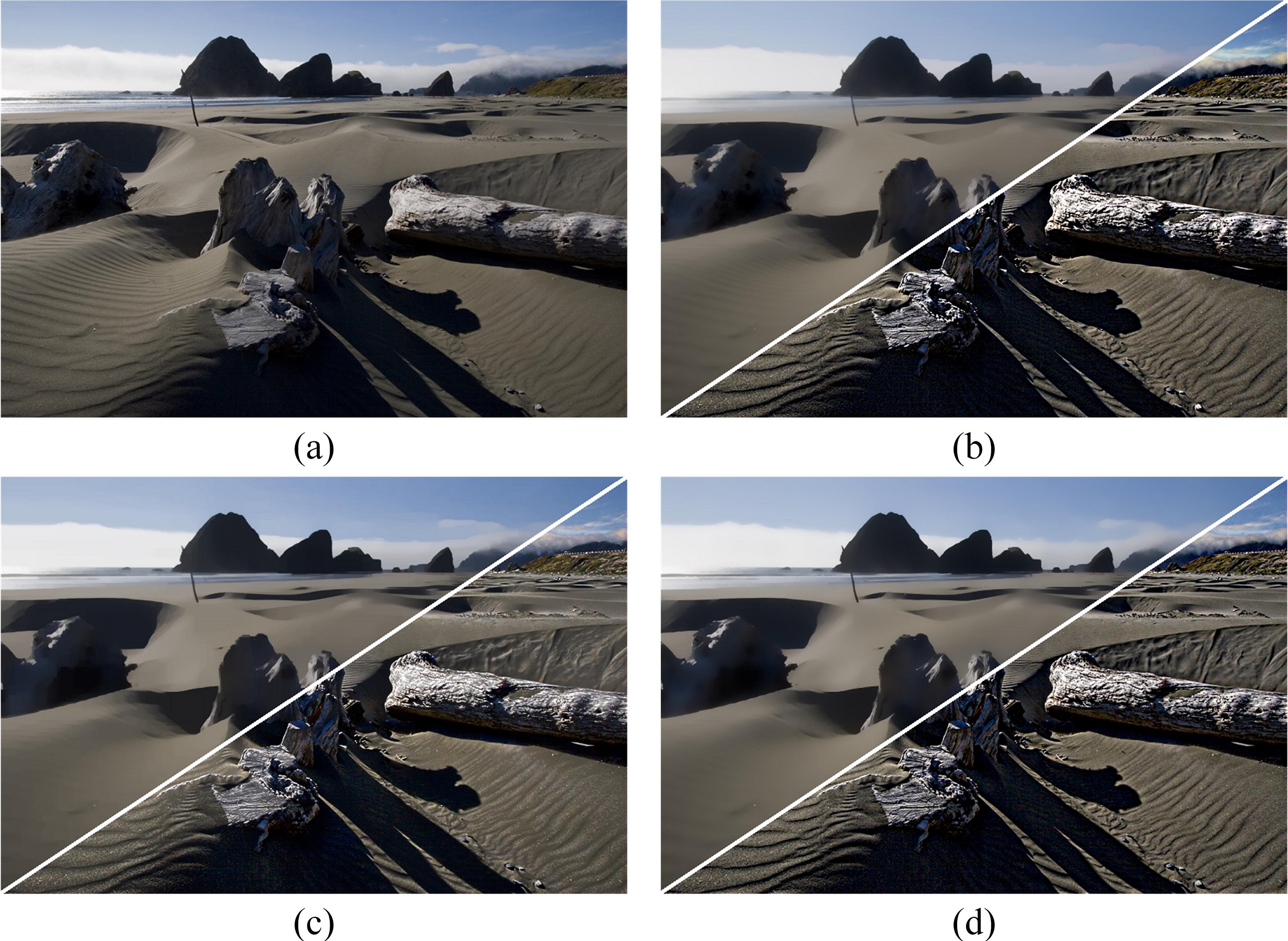}\\
  \caption{(a) Input image. Image smoothing (the upper left part) and detail enhancement (the lower right part) with (b) WLS \cite{farbman2008edge}, time cost is $3.11$ seconds, (c) FGS \cite{min2014fast}, time cost is $0.047$ seconds, (d) our SG-WLS with $r=1, \tau=1$, time cost is $0.14$ seconds. Our SG-WLS is over $20\times$ faster than WLS \cite{farbman2008edge} but can overcome the limitation of FGS \cite{min2014fast}. Zoom in for better visual comparison.}\vspace{-1.5em}\label{FigCover}
\end{figure}

There are also EPS methods based on \emph{global filters} \cite{farbman2008edge, lischinski2006interactive, xu2011image, xu2012structure}. These methods seek a globally optimal solution to the objective function. The objective function usually contains a data constraint term and a prior smoothness term and needs to be solved globally in a principled manner. Global filters can overcome the limitations of local filters such as gradient reversals and halo artifacts \cite{farbman2008edge}. However, most of global filters are time-consuming and some of them are also memory-consuming. The Weight Least Squares (WLS) model has been widely used in image processing and computer graphics \cite{an2008appprop, bhat2010gradientshop, farbman2008edge, lischinski2006interactive}. The solution to the model typically needs to solve a large linear system which is both time- and memory-consuming.

In this paper, we present a new approximation to the WLS model which is denoted as Semi-Global Weighted Least Squares (SG-WLS). {\it The main contributions of this paper are as follows}.

\noindent
~ $-$ Our SG-WLS can achieve close performance to the original WLS model in several challenging tasks. Yet, our SG-WLS is $\sim20$ times faster than the original WLS model. For an image of $M\times N$, the memory cost of SG-WLS is at most at the magnitude of $\max\{\frac{1}{M}, \frac{1}{N}\}$ of that of the original WLS model.
%

\noindent
~ $-$ The 1D filters in our SG-WLS can take two-dimensional neighborhood information into account each time due to the newly proposed neighborhood construction. This is different from previous methods \cite{gastal2011domain, pham2005separable, yang2012recursive, min2014fast} that can only consider neighbors in one dimension each time. Our neighborhood construction thus overcomes the limitation of previous methods and shows improved results in several applications.
%

\noindent
~ $-$ We propose a fast solution to each 1D filter in our SG-WLS, which is denoted as \emph{r-band LU decomposition}, to handle a more general and larger neighborhood system other than the 4-connected/8-connected neighborhood system. Such a generalization can achieve better performance than the 4-connected/8-connected neighborhood system in some applications such as guided depth upsampling.

{\bf Background}
%
The WLS model is a fundamental optimization framework that has been widely used \cite{an2008appprop, bhat2010gradientshop, farbman2008edge, lischinski2006interactive}. Given a target image $F$ to be filtered and a guidance image $G$, the formulation of WLS is defined as:
\begin{equation}\label{EqWLSFormulation}
\small
    E(U) = \sum\limits_{i\in \Omega}(U_i - F_i)^2 + \lambda\sum\limits_{i\in \Omega}\sum\limits_{j\in N(i)}\omega_{i,j}(U_i - U_j)^2
\end{equation}
where $\Omega$ represents the set of all the coordinates. $\lambda$ is a parameter that balances the data term and smoothness term. A larger $\lambda$ results in a larger smoothing effect on $F$. $N(i)$ is the neighborhood of the pixel with coordinate $i$ which is a square patch of $(2r + 1)\times (2r + 1)$ centered at $i$. $\omega_{i,j}$ is the guidance weight based on $G$. Based on different applications which we will detail in Sec.~\ref{SecExperiments}, we adopt two types of guidance weight for different applications:
\begin{equation}\label{EqGuidanceWeight}
\small
\begin{split}
    &\omega_{i,j} = \omega_{i,j}^{frac} = \frac{1}{|i - j|^{\alpha_s} + \varepsilon}\cdot \frac{1}{|G_i - G_j|^{\alpha_r} + \varepsilon}\\
    &\omega_{i,j} = \omega_{i,j}^{exp} = \exp\left(-\frac{|i - j|^2}{2\sigma_s^2}\right) \cdot \exp\left(-\frac{|G_i - G_j|^2}{2\sigma_r^2}\right)
\end{split}
\end{equation}
where $\alpha_s$, $\alpha_r$, $\sigma_s$ and $\sigma_r$ are constants defined by the user. $\varepsilon$ is a small constant that prevents division by zero in areas where $G$ is constant. In this paper, we set $\varepsilon = 0.0001$.

The unique minimum of Eq.~(\ref{EqWLSFormulation}) can be obtained by solving the following large linear system:
\begin{equation}\label{EqLargeLinearSystem}
\small
   A\cdot U = F
\end{equation}
here $U$ and $F$ are vector representations of two-dimensional images. If an image is of size $M\times N$, then its vector representation is a $S\times1$ vector where $S=M\times N$. $A$ is a $S\times S$ matrix and is defined as:
\begin{eqnarray}\label{EqLaplacianMatrix}
\small
    A_{i,j}=\left\{
    \begin{array}{l}
    1 + \lambda\sum_{j\in N(i)}\omega_{i,j} \ \ \ \text{for} \ i=j\\
    -\lambda\omega_{i,j}\ \ \ \ \ \ \ \ \ \ \ \ \ \ \ \ \ \ \ \ \ \ \text{for} \ j\in N(i)\\
    0\ \ \ \ \ \ \ \ \ \ \ \ \ \ \ \ \ \ \ \ \ \ \ \ \ \ \ \  \ \  \ \ \text{otherwise}
    \end{array} \right.
\end{eqnarray}

There are two challenging issues in solving Eq.~(\ref{EqLargeLinearSystem}): (I) Large memory cost. Despite the memory cost of the solver, the memory cost of storing $A$ in Eq.~(\ref{EqLaplacianMatrix}) is $\mathcal{O}(MNr^2)$. As $r$ becomes larger, the memory cost increases very fast. (II) Large time cost. Solving Eq.~(\ref{EqLargeLinearSystem}) is an inverse problem which is time-consuming since Eq.~(\ref{EqLargeLinearSystem}) is very large. Note that the time cost also increases as $r$ becomes larger.

Eq.~(\ref{EqLargeLinearSystem}) can be directly solved by modern linear solvers such as Preconditioned Conjugate Gradient (PCG) \cite{koutis2011combinatorial, krishnan2013efficient}. However, the convergence of PCG strongly depends on a good choice of the preconditioner \cite{lischinski2006interactive}. Besides, the memory cost is not reduced. To solve the above two issues, various approximate methods have been proposed in the literature. Barron et al. \cite{barron2016fast} proposed to first project the original image into a bilateral space. Then a much smaller linear system is solved with PCG \cite{shewchuk1994introduction}. The output is projected back into the original image space which is the final filtered result. The method in \cite{barron2015fast} shares the similar idea. Both of their methods are post-processed by the domain transform filtering \cite{gastal2011domain} to smooth out the blocky artifacts introduced by the bilateral grid. Xu et al. \cite{xu2009efficient} proposed to first cluster the image in an affinity space with \emph{kd} tree and then solve another linear system with the clustered image. In summary, the key idea of these methods is to reduce the dimension of either the large matrix ($``A"$ in Eq.~(\ref{EqLargeLinearSystem})) or the input image. Thus the inverse operation of the very large system can be reduced to the matrix inverse of much smaller matrixes which can reduce both time and memory cost.

The Fast Global Smoother (FGS) proposed by Min et al. \cite{min2014fast} is closely related to our work. FGS divides the WLS model into a sequence of subsystems in each row and column which is different from the previous methods \cite{an2008appprop, barron2015fast, barron2016fast, xu2009efficient}. As FGS is locally global in each row and column, we denote it as a \emph{semi-global} method. Our work is also a semi-global method. It is different from FGS \cite{min2014fast} in the following two aspects: (I) Our construction of subsystems is different from that in FGS. FGS only considers neighbors in one dimension each time (row direction or column direction) which has largely destroyed the two-dimensional neighborhood system in the original model. This can cause noticeable artifacts as illustrated in Fig.~\ref{FigCover}(c). On the contrary, our method can take two-dimensional neighborhood information into account. Thus, our method can overcome the limitation of FGS and achieves close performance to the original WLS model as illustrated in Fig.~\ref{FigCover}(d). (II) The FGS only adopts the 4-connected/8-connected neighborhood system while our method can handle a more general $(2r+1)\times(2r+1)$ neighborhood system thanks to the proposed fast solution. We show that such a generalization makes our method achieve better performance than the 4-connected/8-connected one in some applications such as guided depth upsampling \cite{ferstl2013image, li2016fast}.

\begin{figure*}[t!]
\centering
  \includegraphics[width=0.959\linewidth]{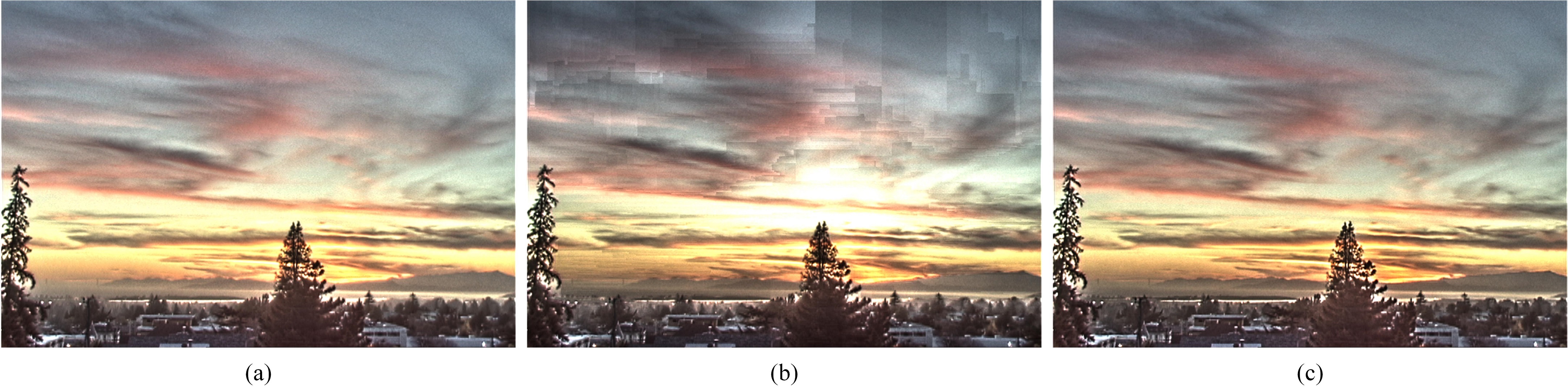}\\
  \caption{Visual comparison of HDR tone mapping. Results of (a) WLS \cite{farbman2008edge}, (b) FGS \cite{min2014fast} and (c) our SG-WLS with $r=1, \tau=1$. Blocky artifacts are noticeable in (b).}
  \vspace{-.215em}
  \label{FigShortageOfGFS}
\end{figure*}

\section{Semi-Global Weighted Least Squares}
\subsection{Neighborhood Construction of Subsystems}
\label{SecNeighborhoodConstruction}
\begin{figure}[t!]
\centering
  \includegraphics[width=0.99\linewidth]{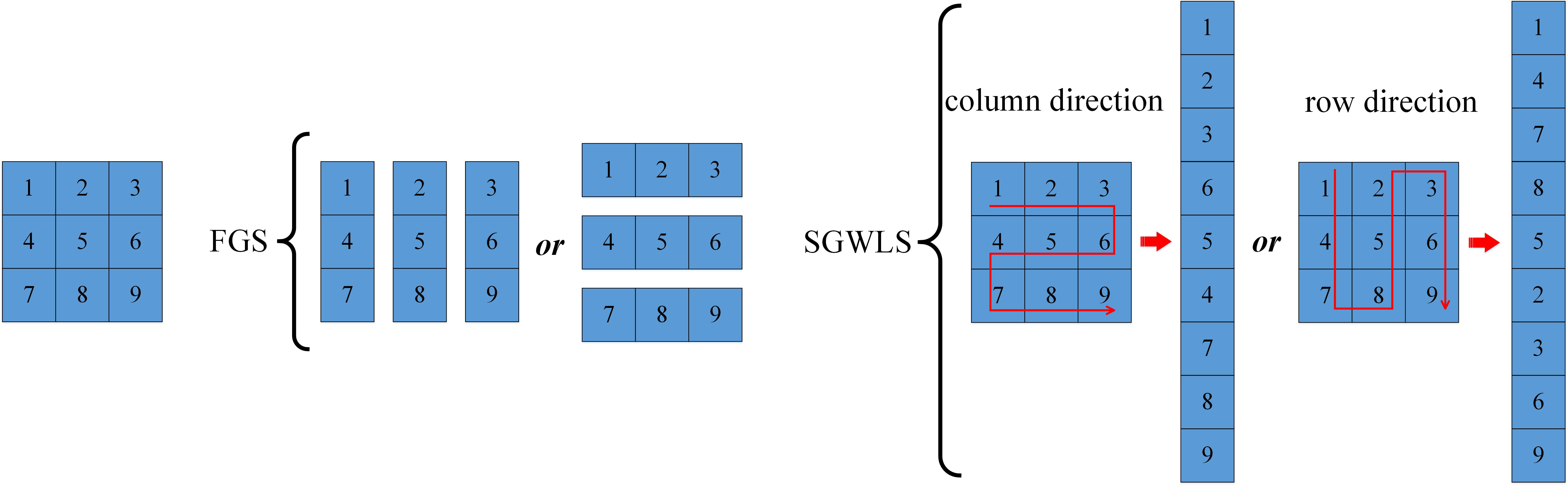}\\
  \caption{Illustration of neighborhood construction in subsystems of FGS \cite{min2014fast} and our SG-WLS.}\vspace{-1.5em}\label{FigNeighborComparison}
\end{figure}

Like many previous separate filters \cite{gastal2011domain, pham2005separable, yang2012recursive}, the FGS proposed by Min et al. \cite{min2014fast} separates a two-dimensional filtering process into an iterative one-dimensional filtering process. FGS is fast and can process a 1-megapixel RGB image in $0.1$ seconds on a standard desktop. However, the main limitation of this separation is that only neighborhood in one dimension is considered each time. In such cases, the original two-dimensional neighbors has been largely  destroyed as illustrated in Fig.~\ref{FigNeighborComparison}. In some applications, such limitation can cause noticeable artifacts. We show some examples of image detail enhancement and HDR tone mapping obtained by FGS in Fig.~\ref{FigCover}(c) and Fig.~\ref{FigShortageOfGFS}(b) respectively.

In this section, we show a new type of neighborhood construction that can handle neighbors in two dimensions each time within one-dimensional separate filters. For a pixel at row $s$ and column $t$ in image $I$ of size $M\times N$, we denote it as $I_{(s,t)}$. For a WLS model with a $(2r+1)\times(2r+1)$ neighborhood system, we first extract $2r+1$ columns around the $k$th ($k=r+1,\cdots,N-r$) column in the image, which are denoted as $[I_{(*,k-r)}, I_{(*,k-r+1)},\cdots,I_{(*,k)},\cdots,I_{(*,k+r-1)}, I_{(*,k+r)}]$. Here $I_{(*,k)}$ denotes all the pixels in the $k$th column. If arranged in their original order, these $2r+1$ columns totally have $M$ row vectors of size $2r+1$. Then for each $i$th ($i=1,\cdots,M$) row vector, if $i$ is even, we reverse the order of the $i$th row vector. Finally, these row vectors are connected head to end to form a $(2r+1)M\times 1$ column vector. We denote this process as neighborhood construction along column direction. It can also be performed along row direction for each $k$th ($k=r+1,\cdots,M-r$) row. In this way, the formed 1D vector is a $(2r+1)N\times 1$ one. An example of $r=1$ is illustrated in Fig.~\ref{FigNeighborComparison}.

Now we show how the above neighborhood construction can handle neighbors in two dimensions. First, note that for any $2r+1$ neighboring pixels in the formed 1D vector, they are also neighbors in the original $(2r+1)\times(2r+1)$ neighborhood system. Then take the neighborhood construction along column direction for example, the final 1D vector also contains pixels from neighboring $r$ columns on each side of the current column, which also contains neighbors in the row direction. In this way, the final vector can handle neighbors from both row and column directions. As illustrated in Fig.~\ref{FigNeighborComparison}, pixel $``4"$ and pixel $``6"$ are neighbors of pixel $``5"$ in row direction in the original neighborhood system. In the neighborhood construction along column direction, pixel $``4"$ and pixel $``6"$ are still neighbors of pixel $``5"$. This newly designed neighborhood construction can well overcome the limitation of previous separate filters that only handle neighbors in one dimension \cite{gastal2011domain, pham2005separable, yang2012recursive}. Fig.~\ref{FigCover}(d) and Fig.~\ref{FigShortageOfGFS}(c) show results obtained with our newly designed neighborhood construction. Our results are indistinguishable to the ones obtained with WLS \cite{farbman2008edge} and well overcome the limitation of FGS \cite{min2014fast}. Details of our method will be described in Sec.~\ref{SecSG-WLSArchitecture}.

\subsection{Fast and Exact Solution to Subsystems}
\label{SecSubsystemSolution}
Since any neighboring $2r+1$ pixels in the 1D vector formed in the neighborhood construction step are neighboring pixels in the original two-dimensional neighborhood system, we can solve another WLS model using this 1D vector with a $2r+1$ neighborhood system. Similarly, for each formed 1D vector, we can obtain a much smaller linear system as follows:
\begin{equation}\label{EqSubLinearSystem}
\small
    A_s\cdot u=f
\end{equation}
where $f$ is the formed 1D vector in the neighborhood construction step. $A_s$ is formed in a way similar to Eq.~(\ref{EqLaplacianMatrix}). Eq.~(\ref{EqSubLinearSystem}) is denoted as a subsystem of our method. Note that the neighborhood system $N(i)$ here is one-dimensional $2r+1$ pixels. To be explicit, $A_s$ has the following structure:
\begin{equation*}\label{EqStructureOfA}
\footnotesize
\hspace{-.51cm}
A_s=
\begin{bmatrix}
\begin{smallmatrix}
   a_{1}   & b_{1,1}   & b_{1,2}   & \cdots  & b_{1,r}   &           &           &           &           &       \\

   c_{1,1} & a_{2}     & b_{2,1}   & \cdots  & b_{2,r-1} & b_{2,r}   &           &           &           &        \\

   c_{1,2} & c_{2,1}   & a_{3}     & \cdots  & b_{3,r-2} & b_{3,r-1} & b_{3,r}   &           &           &         \\

   \vdots  & \vdots    & \vdots    & \ddots  & \ddots    & \ddots    & \ddots    & \ddots    &           &         \\

   c_{1,r} & c_{2,r-1} & c_{3,r-2} & \ddots  & \ddots    & \ddots    & \ddots    & \ddots    & \ddots    &          \\

           & c_{2,r}   & c_{3,r-1} & \ddots  & \ddots    & a_{s-r}   & b_{s-r,1} & b_{s-r,2} & \cdots    & b_{s-r,r} \\

           &           & c_{3,r}   & \ddots  & \ddots    & b_{s-r,1} & \ddots    & \ddots    & \ddots    & \vdots    \\

           &           &           & \ddots  & \ddots    & b_{s-r,2} & \ddots    & a_{s-2}   & b_{s-2,1} & b_{s-2,2}  \\

           &           &           &         & \ddots    & \vdots    & \ddots    & c_{s-2,1} & a_{s-1}   & b_{s-1,1}  \\
           &           &           &         &           & c_{s-r,r} & \cdots    & c_{s-2,2} & c_{s-1,1} & a_{s}       \\
\end{smallmatrix}
\end{bmatrix}
\end{equation*}
here $s=(2r+1)M$ for neighborhood construction along column direction and $s=(2r+1)N$ for neighborhood construction along row direction.

Matrix $A_s$ in Eq.~(\ref{EqSubLinearSystem}) is positive definite and it is also a diagonal matrix with bandwidth of $r$. For $r=1$, solving Eq.~(\ref{EqSubLinearSystem}) has been well studied and has classical solution such as LU decomposition \cite{golub2012matrix}. However, here we need to solve a more general case with $r\geq1$ which is seldom studied. In this section, we propose a solution to Eq.~(\ref{EqSubLinearSystem}) to handle the case for any $r\geq1$, which is denoted as \emph{$r$-band LU decomposition}. Moreover, such a decomposition can be completed efficiently. To be more explicit, this process can be formulated as follows:
\begin{equation}\label{EqLUDecomposition}
\small
    A_s = P\cdot Q
\end{equation}
\begin{equation}\label{EqLowerEq}
\small
    P\cdot y = f
\end{equation}
\begin{equation}\label{EqUpperEq}
\small
    Q\cdot u = y
\end{equation}

Assuming $P$ and $Q$ have the following structures:
\begin{equation*}\label{EqSturctureOfL}
\small
P=
\begin{bmatrix}
\begin{smallmatrix}
\alpha_{1}   &                &                &           &           &          &          &          &          &      &\\

\gamma_{1,1} & \alpha_{2}     &                &           &           &          &          &          &          &      &\\

\gamma_{1,2} & \gamma_{2,1}   & \alpha_{3}     &           &           &          &          &          &          &      &\\

\vdots       & \vdots         & \vdots         &\ddots     &           &          &          &          &          &      &\\

\gamma_{1,r} & \gamma_{2,r-1} & \gamma_{3,r-2} &\ddots     &\ddots     &          &          &          &          &      &\\

             & \gamma_{2,r}   & \gamma_{3,r-1} & \ddots    &\ddots     & \alpha_{s-r}   &          &          &          &\\

             &                & \gamma_{3,r}   & \ddots    &\ddots     & \gamma_{s-r,1}   & \ddots   &          &         &\\

             &                &                & \ddots    &\ddots     & \vdots   & \ddots   &\ddots  &      &\\

             &                &                &           &\ddots     & \vdots   & \ddots   &\ddots  &  \alpha_{s-1}     &\\

             &      &      &            &       &\gamma_{s-r,r} & \cdots & \gamma_{s-2,2} & \gamma_{s-1,1} & \alpha_{s}\\
\end{smallmatrix}
\end{bmatrix}
\end{equation*}
\begin{equation*}\label{EqSturctureOfL}
\small
Q=
\begin{bmatrix}
\begin{smallmatrix}
1 & \beta_{1,1} & \beta_{1,2}  & \cdots & \beta_{1,r}   &               &           \\

  & 1           & \beta_{2,1} & \cdots & \beta_{2,r-1} & \beta_{1,r}   &           \\

  &             & 1           & \cdots & \beta_{3,r-2} & \beta_{2,r-1} & \beta_{1,r}\\

  &             &             & \ddots & \ddots        & \ddots        & \ddots      & \ddots \\

  &             &             &        & 1        & \beta_{s-r, 1}       & \cdots      & \cdots  & \beta_{s-r,r}\\

  &             &             &        &          & \ddots               & \ddots      & \ddots  & \vdots\\

  &             &             &        &          &                      & 1           & \beta_{s-2, 1}  & \beta_{s-2, 2}\\

  &             &             &        &          &                      &             &    1            & \beta_{s-1, 1}\\

  &             &             &        &          &                      &           &    & 1\\
\end{smallmatrix}
\end{bmatrix}
\end{equation*}

According to Eq.~(\ref{EqLUDecomposition}), for $i=1,\cdots,r$, we have:
\begin{equation}\label{EqLUDecompositionEq1}
\small
\begin{array}{l}
\alpha_1=a_1,\ \gamma_{1,i}=c_{1,i},\ \beta_{1,i}=b_{1,i}/\alpha_{1}
\end{array}
\end{equation}

For $k=2,\cdots,r$ and $i=1,\cdots,r-k+1$, we have:
\begin{eqnarray}\label{EqLUDecompositionEq2}
\small
\begin{array}{l}
\alpha_k\ \ = a_k - \sum_{t=1}^{k-1}\gamma_{k-t,t}\beta_{k-t,t},\\
\gamma_{k,i}=c_{k,i}-\sum_{t=1}^{k-1}\gamma_{k-t,i+t}\beta_{k-t,t},\\
\beta_{k,i} = \frac{1}{\alpha_k}(b_{k,i} - \sum_{t=1}^{k-1}\beta_{k-t,i+t}\gamma_{k-t,t})
\end{array}
\end{eqnarray}

For $k=3,\cdots,r$ and $i=r-k+2,\cdots,r-1$, we have:
\begin{eqnarray}\label{EqLUDecompositionEq3}
\small
\begin{array}{l}
\gamma_{k,i}=c_{k,i}-\sum_{t=1}^{r-i}\gamma_{k-t,i+t}\beta_{k-t,t},\\
\beta_{k,i} = \frac{1}{\alpha_k}(b_{k,i} - \sum_{t=1}^{r-i}\beta_{k-t,i+t}\gamma_{k-t,t})
\end{array}
\end{eqnarray}

For $k=r+1,\cdots,s$, we have:
\begin{eqnarray}\label{EqLUDecompositionEq4}
\small
\begin{array}{l}
\alpha_k\ \ = a_k - \sum_{t=1}^{r}\gamma_{k-t,t}\beta_{k-t,t},\\
\end{array}
\end{eqnarray}

For $k=r+1,\cdots,s-1$ and $i=1,\cdots,\min\{r-1,s-k\}$, we have:
\begin{eqnarray}\label{EqLUDecompositionEq5}
\small
\begin{array}{l}
\gamma_{k,i}=c_{k,i}-\sum_{t=1}^{r-i}\gamma_{k-t,i+t}\beta_{k-t,t},\\
\beta_{k,i} = \frac{1}{\alpha_k}(b_{k,i} - \sum_{t=1}^{r-i}\beta_{k-t,i+t}\gamma_{k-t,t})
\end{array}
\end{eqnarray}

For $k=2,\cdots,s-r$, we have:
\begin{eqnarray}\label{EqLUDecompositionEq6}
\small
\begin{array}{l}
\gamma_{k,r}=c_{k,r},\beta_{k,r} = \frac{b_{k,r}}{\alpha_k}
\end{array}
\end{eqnarray}

Eqs.~(\ref{EqLUDecompositionEq1})$\sim$(\ref{EqLUDecompositionEq6}) are the r-band LU decomposition in Eq.~(\ref{EqLUDecomposition}). When this is completed, we can solve Eq.~(\ref{EqSubLinearSystem}) through Eq.~(\ref{EqLowerEq}) and Eq.~(\ref{EqUpperEq}). When solving Eq.~(\ref{EqLowerEq}), we first have:
\begin{equation}\label{EqLowerEq1}
\small
\begin{array}{l}
    y_{1} = \frac{f_1}{\alpha_1}
\end{array}
\end{equation}

For $k=2,\cdots,r$, we have:
\begin{eqnarray}\label{EqLowerEq2}
\small
\begin{array}{l}
    y_{k} = \frac{1}{\alpha_k}(f_k - \sum_{t=1}^{k-1}\gamma_{t,k-t}y_t)
\end{array}
\end{eqnarray}

For $k=r+1,\cdots,s$, we have:
\begin{eqnarray}\label{EqLowerEq3}
\small
\begin{array}{l}
    y_{k} = \frac{1}{\alpha_k}(f_k - \sum_{t=1}^{r}\gamma_{k-t,t}y_{k-t})
\end{array}
\end{eqnarray}

When solving Eq.~(\ref{EqUpperEq}), we first have:
\begin{equation}\label{EqUpperEq1}
\small
\begin{array}{l}
u_s=y_s
\end{array}
\end{equation}

For $k=s-1,\cdots,s-r+1$, we have:
\begin{equation}\label{EqUpperEq2}
\small
\begin{array}{l}
u_k=y_k-\sum_{t=1}^{s-k}\beta_{k,t}u_{k+t}
\end{array}
\end{equation}

For $k=s-r,\cdots,1$, we have:
\begin{equation}\label{EqUpperEq3}
\small
\begin{array}{l}
u_k=y_k-\sum_{t=1}^{r}\beta_{k,t}u_{k+t}
\end{array}
\end{equation}

The subsystem in Eq.~(\ref{EqSubLinearSystem}) can be solved exactly through Eqs.~(\ref{EqLUDecompositionEq1})$\sim$(\ref{EqUpperEq3}). Note that for $r=1$, only Eqs.~(\ref{EqLUDecompositionEq1}), (\ref{EqLUDecompositionEq4}), (\ref{EqLUDecompositionEq6}), (\ref{EqLowerEq1}), (\ref{EqLowerEq3}), (\ref{EqUpperEq1}) and (\ref{EqUpperEq3}) are needed. For $r=2$, only Eq.~(\ref{EqLUDecompositionEq3}) is not needed.

\subsection{Semi-global Edge-preserving Smoothing}
\label{SecSG-WLSArchitecture}
Based on the subsystems described in Sec.~\ref{SecNeighborhoodConstruction} and Sec.~\ref{SecSubsystemSolution}, we can perform edge-preserving smoothing which can benefit numerous applications \cite{farbman2008edge, min2014fast}. For a $M\times N$ image with a $(2r+1)\times(2r+1)$ two-dimensional neighborhood system, our goal is to divide the original two-dimensional WLS model described in Eq.~(\ref{EqWLSFormulation}) into a sequence of one-dimensional WLS models with a $2r+1$ one-dimensional neighborhood system in Eq.~(\ref{EqSubLinearSystem}).

Our smoothing process contains four steps. The first step is the one-dimensional neighborhood construction described in Sec.~\ref{SecNeighborhoodConstruction}. $2r+1$ columns (along column direction) or rows (along row direction) in the original image are needed each time. This step results in 1D vectors of size $(2r+1)M$ along column direction or $(2r+1)N$ along row direction. Each vector has a $2r+1$ neighborhood system. The second step is solving the linear system with the formed 1D vector which is described in Sec.~\ref{SecSubsystemSolution} . The third step is transforming the solution in the second step into an image patch of $2r+1$ rows/columns. This is a simple inverse operation of the neighborhood construction in the first step. The fourth step is averaging pixel values. This is because one pixel can be involved in several subsystems. These values in different subsystems of the same pixel are averaged as the final output. The above four steps are performed $T$ times along column direction and row direction alternatively to get the final smoothed image. In this paper, we find $T=2\sim4$ is appropriate for most applications. As each subsystem is a globally optimized one while it is only performed in a local region of an image, we call our method as Semi-Global Weighted Least Squares (SG-WLS).

When SG-WLS is used for sparse interpolation such as guided depth upsampling \cite{ferstl2013image, li2016fast} and colorization \cite{levin2004colorization}, it cannot be directly applied to the sparse input data due to the unstable result. Instead, we perform guided sparse interpolation in a way similar to the one in \cite{lang2012practical, li2016fast, min2014fast}. Denote the index map of input as $H$, then the output is computed as:
\begin{equation}\label{EqSparseInterpolation}
\small
    U(m)=\frac{A^{-1}F(m)}{A^{-1}H(m)}, m \in \Omega
\end{equation}

The WLS smoothing is applied to both the input $F$ and the index map $H$. This procedure is approximated with our SG-WLS.

\begin{figure}
\centering
  \includegraphics[width=0.8\linewidth]{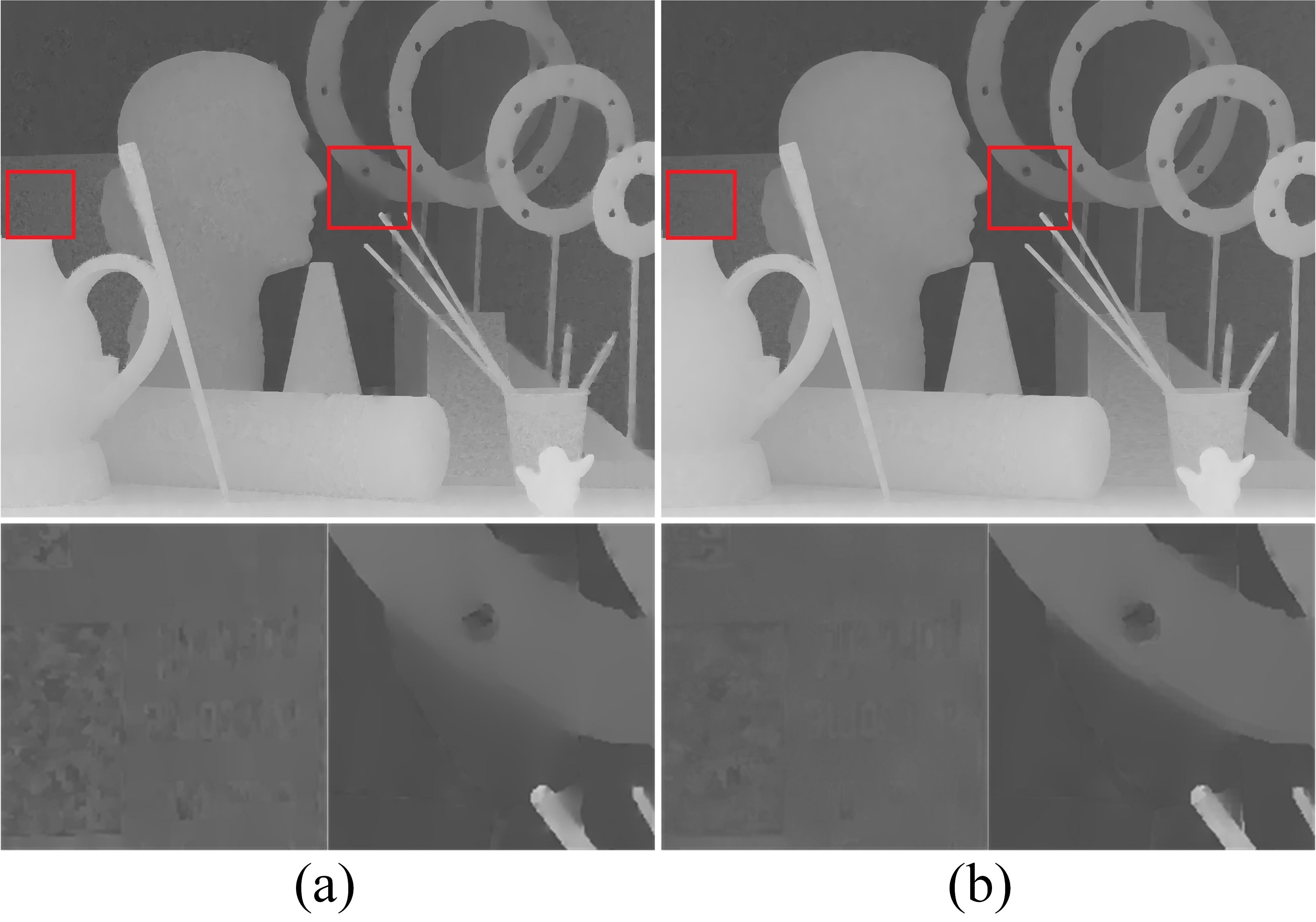}\\
  \caption{Visual comparison of the advantage of using a lager neighborhood system radius. $4\times$ guided upsampling result of our SG-WLS with (a) $r=1$, $\sigma_s=1$, $\sigma_r=3$, $\lambda=900$, the MAD is $2.4$, (b) $r=4$, $\sigma_s=4$, $\sigma_r=3$, $\lambda=200$, the MAD is $1.9$.}\vspace{-0.5em}\label{FigLargerRadiusAdvantage}
\end{figure}

Different from the FGS \cite{min2014fast} that only adopts the 4-connected/8-connected neighborhood system, our SG-WLS focuses on a more general case where the neighborhood system is $(2r+1)\times(2r+1)$. In fact, the 8-connected neighborhood system in the FGS \cite{min2014fast} is a special case of our neighborhood system with $r=1$. In their work, the smoothing strength is increased by enlarging $\lambda$. For our SG-WLS, we show that the smoothing strength can also be increased by enlarging the radius $r$ of the neighborhood system. Particularly, in some cases, enlarging $r$ can achieve the smoothing property that cannot be achieved by simply enlarging $\lambda$. Fig.~\ref{FigLargerRadiusAdvantage} shows one example of guided depth upsampling. When enlarging $\lambda$, depth edges have been blurred while some parts are still noisy as highlighted. However, when we use a larger $r$ but a smaller $\lambda$, depth edges are well preserved while the noise is also smoothed. Note that Yang et al. \cite{yang2014color} also adopted an $11\times11$ neighborhood system other than a 4-connected/8-connected one for guided depth upsampling. This also validates the effectiveness of larger neighborhood systems. Quantitative measurement of Mean Absolute Difference (MAD) between the result and the groundtruth also shows the effectiveness of using larger neighborhood systems.

\begin{figure*}
\centering
  \includegraphics[width=0.872\linewidth]{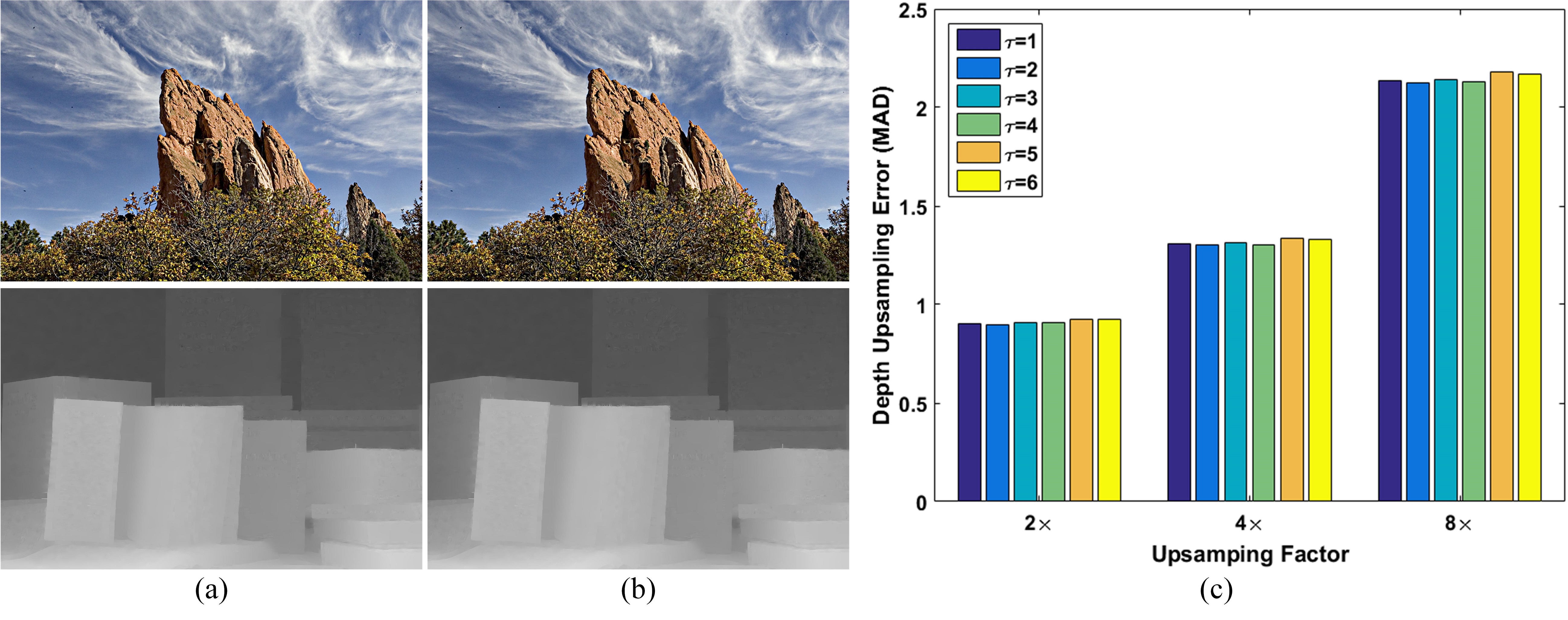}\\
  \caption{Comparison of image detail enhancement and guided depth upsampling using different values of $\tau$. Results are obtained with $r=4$ and (a) $\tau=1$, (b) $\tau=4$. (c) MAD comparison of guided depth upsampling under different values of $\tau$.}\vspace{-1.5em}\label{FigStepEffect}
\end{figure*}

\begin{figure}[h!]
\centering
  \includegraphics[width=0.78\linewidth]{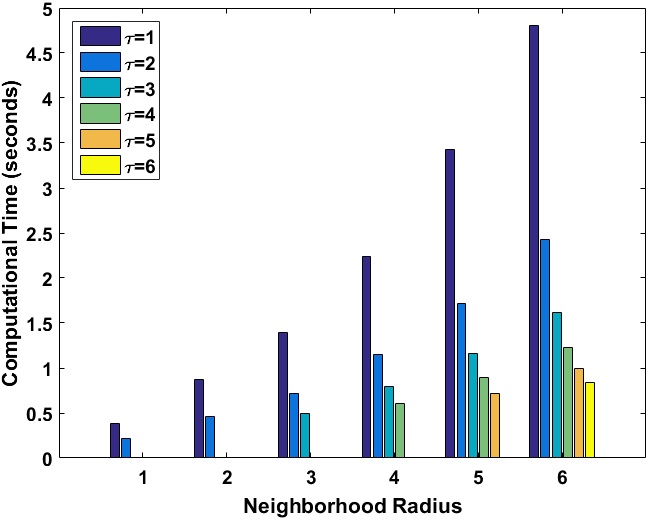}\\
  \caption{Computation time of our SG-WLS for smoothing a 1-megapixel RGB image with different parameter settings.}\vspace{-1.5em}\label{FigDifferentStepTime}
\end{figure}

The neighborhood construction in the first step can be performed in a sliding manner. Take the neighborhood construction along column direction for example, when a 1D vector is formed with $2r+1$ columns centered at the $k$th column, i.e., $[I_{(*,k-r)}, I_{(*,k-r+1)},\cdots,I_{(*,k)},\cdots,I_{(*,k+r-1)}, I_{(*,k+r)}]$. Then the next 1D vector is formed with $2r+1$ columns centered at the $(k+1)$th column. However, we find this kind of neighborhood construction is redundant. We can slide it with a step $\tau$, i.e., for the above case, the next 1D vector can formed with $2r+1$ columns centered at the $(k+\tau)$th column. According to our experimental results, for $\tau\leq r$, our SG-WLS has similar performance to that of $\tau=1$ but with smaller computational cost. Fig.~\ref{FigStepEffect} shows an example of detail enhancement and guided depth upsampling. The MAD of guided depth upsampling results in Fig.~\ref{FigStepEffect}(c) also shows increasing $\tau$ seldom decreases the performance. Fig.~\ref{FigDifferentStepTime} shows computation time comparison among different values of $\tau$. As shown in Fig.~\ref{FigDifferentStepTime}, increasing $\tau$ can greatly reduce the computational cost.

\begin{figure*}[t!]
\centering
  \includegraphics[width=0.9\linewidth]{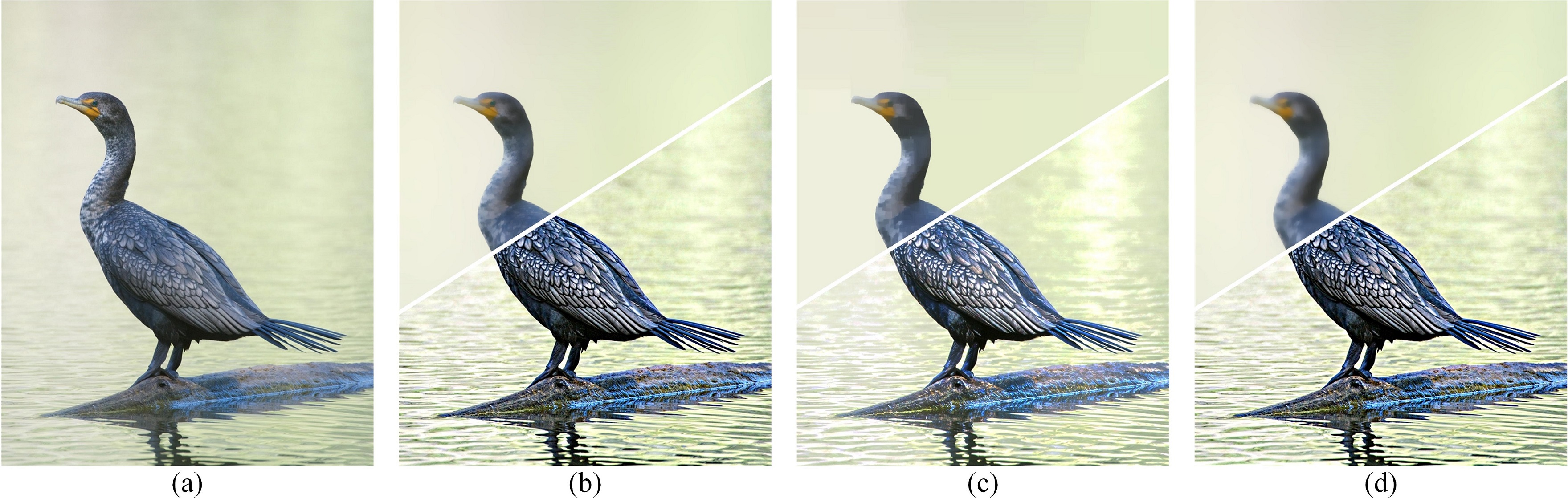}\\
  \caption{Image detail enhancement comparison. (a) Input image. Results of (b) WLS \cite{farbman2008edge}, (c) FGS \cite{min2014fast} and (d) our SG-WLS. The result in (b) is pale while the ones in (a) and (c) are more colorful. }\vspace{-1.5em}\label{FigDetailEnhancement}
\end{figure*}

{\bf Computation Complexity}
%
%
\begin{table}[b!]
\caption{Computation time of different methods for smoothing a 1-megapixel RGB image.}\label{TabDifferentMethodsTime}
\resizebox{1\linewidth}{!}
{
\begin{tabular}{c|c|c|c}
  \Xhline{1.2pt}
  Method & WLS \cite{farbman2008edge} & FGS \cite{min2014fast} & Ours ($r=1,\tau=1$) \\
  \hline
  Time (seconds) & 8.02 & 0.1 & 0.35 \\
  \Xhline{1.2pt}
\end{tabular}
}
\end{table}
%
Assume that the image is of size $M\times N$. The neighborhood system is $(2r+1)\times(2r+1)$ in the original two-dimensional WLS model. Thus, the neighborhood system of our SG-WLS is $2r+1$. Despite the few elements near the boundary, when solving a subsystem in Eq.~(\ref{EqSubLinearSystem}), there are $r$ times multiplication for each $\alpha_k$ and $\gamma_{k,i}$, $r+1$ times multiplication for each $\beta_{k,i}$, $r+1$ times addition for $\alpha_k$, $\gamma_{k,i}$ and $\beta_{k,i}$. Thus, Eqs.~(\ref{EqLUDecompositionEq1})$\sim$(\ref{EqLUDecompositionEq6}) require $\mathcal{O}(sr+sr^2+sr(r+1))$ times multiplication operations and $\mathcal{O}(s(r+1)+2sr(r+1))$ times addition operations. Since $s=(2r+1)M$ along column direction and $s=(2r+1)N$ along row direction, the computational complexity is $\mathcal{O}(Mr^3)$ or $\mathcal{O}(Nr^3)$ for both multiplication and addition. Similarly, solving Eq.~(\ref{EqLowerEq}) and Eq.~(\ref{EqUpperEq}) through Eqs.~(\ref{EqLowerEq1})$\sim$(\ref{EqUpperEq3}) requires $\mathcal{O}(Mr^2)$ or $\mathcal{O}(Nr^2)$ multiplication and addition operations. Thus, the computational complexity of solving a subsystem in Eq.~(\ref{EqSubLinearSystem}) is $\mathcal{O}(Mr^3)$ or $\mathcal{O}(Nr^3)$. There are total $\frac{N}{\tau}$ subsystems when applying SG-WLS along column direction and $\frac{M}{\tau}$ subsystems along row direction. Thus, when applied $T$ times, the final computational complexity of SG-WLS is $\mathcal{O}(\frac{T}{\tau}MNr^3)$. Fig.~\ref{FigDifferentStepTime} shows the computation time of different $r$ and $\tau$ for filtering a 1-megapixel RGB image on a computer with an Intel i7 3.40GHz CPU and 8GB memory. The iteration number is fixed with $T=4$. In particular, we compare the computation time of our SG-WLS of $r=1,\tau=1$ with that of the WLS model \cite{farbman2008edge} and FGS \cite{min2014fast} in Table~\ref{TabDifferentMethodsTime}. Our method is over $20\times$ faster than the WLS model \cite{farbman2008edge}. Although our SG-WLS is a little slower than FGS \cite{min2014fast}, it can overcome several limitations of FGS \cite{min2014fast} and achieves close performance to the WLS model \cite{farbman2008edge}.

Storing matrixes $A_s, P, Q$ in Eq.~(\ref{EqLUDecomposition}) is the main memory cost of our SG-WLS which is $\mathcal{O}(sr)$. Since the memory cost of Eqs.~(\ref{EqLUDecompositionEq1})$\sim$(\ref{EqUpperEq3}) is $\mathcal{O}(s)$, the final memory cost of our SG-WLS is $\mathcal{O}(sr+s)$ which is $\mathcal{O}(Mr^2)$ or $\mathcal{O}(Nr^2)$. Note that despite the memory cost of the solver which may be different among different solvers, the memory cost of storing the matrix $A$ in the original WLS model in Eq.~(\ref{EqLargeLinearSystem}) is $\mathcal{O}(MNr^2)$. Thus, the memory cost of our SG-WLS is at most at the magnitude of $max\{\frac{1}{M}, \frac{1}{N}\}$ of that of the WLS model.

\section{Applications and Experimental Results}
\label{SecExperiments}

We test our SG-WLS on four applications including image detail enhancement, HDR tone mapping, guided depth upsampling and image colorization. For the first two applications which represent applications with dense input data, we adopt $\omega_{i,j}=\omega_{i,j}^{frac}$ in Eq.~(\ref{EqGuidanceWeight}). For the rest two applications, we adopt $\omega_{i,j}=\omega_{i,j}^{exp}$ in Eq.~(\ref{EqGuidanceWeight}). These two applications represent applications with sparse input data which can be regular (guided depth upsampling) or irregular (image colorization). For more experimental results, please refer to our supplementary materials.

\noindent \textbf{Image detail enhancement} aims at enhancing the details of an image while avoiding artifacts such as gradient reversals and halos \cite{farbman2008edge, he2013guided}. In our experiments, an image is decomposed into a base layer and a detail layer through edge-preserving smoothing such as FGS \cite{min2014fast} and WLS \cite{farbman2008edge}. Parameters of our SG-WLS are set as follows: $r=1, \tau=1, \alpha_s=\alpha_r=1.2, \lambda=900$. Similar to the result in Fig.~\ref{FigCover}, we illustrate another example of our detail enhancement results and comparison with results of other methods in Fig.~\ref{FigDetailEnhancement}. All the results are obtained by enlarging $3$ times of their corresponding detail layer. Visual comparison shows that our SG-WLS can achieve close performance to the WLS \cite{farbman2008edge}. Note that the result of FGS appears pale while results of our SG-WLS and WLS \cite{farbman2008edge} are more colorful. Please zoom in for better visual comparison.

\begin{figure*}[h!]
\centering
  \includegraphics[width=1\linewidth]{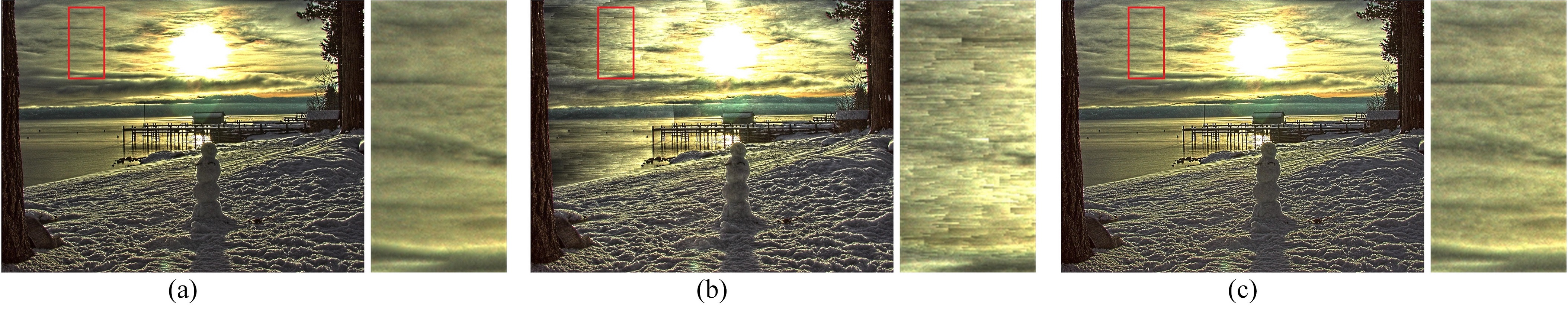}\\
  \caption{HDR tone mapping results of (a) WLS \cite{farbman2008edge}, (b) FGS \cite{min2014fast} and (c) our SG-WLS. Regions in red boxes are highlighted. The result in (b) shows noticeable blocky artifacts.}\label{FigHDRToneMapping}
\end{figure*}

\noindent \textbf{HDR tone mapping} is another application that needs edge-preserving smoothing. Based on the multi-scale tone mapping framework proposed by Farbman et al. \cite{farbman2008edge} \footnote{The source code can be downloaded here \url{http://www.cs.huji.ac.il/~danix/epd/}}, the input image is decomposed into a base layer and three detail layers. The base layer is nonlinearly mapped to a low dynamic range and is re-combined with detail layers. Filters are applied to the logarithmic HDR images. Parameters of our SG-WLS are set as follows: $r=1, \tau=1, \alpha_s=\alpha_r=1.2$ and $\lambda=5/40/320$ for the first/second/third detail layer. Results are illustrate in Fig.~\ref{FigShortageOfGFS} as well as Fig.~\ref{FigHDRToneMapping}. There are noticeable blocky artifacts in the results of FGS \cite{min2014fast} while our SG-WLS can well overcome this limitation and shows close performance to WLS \cite{farbman2008edge}.

\begin{figure*}[h!]
\centering
  \includegraphics[width=1\linewidth]{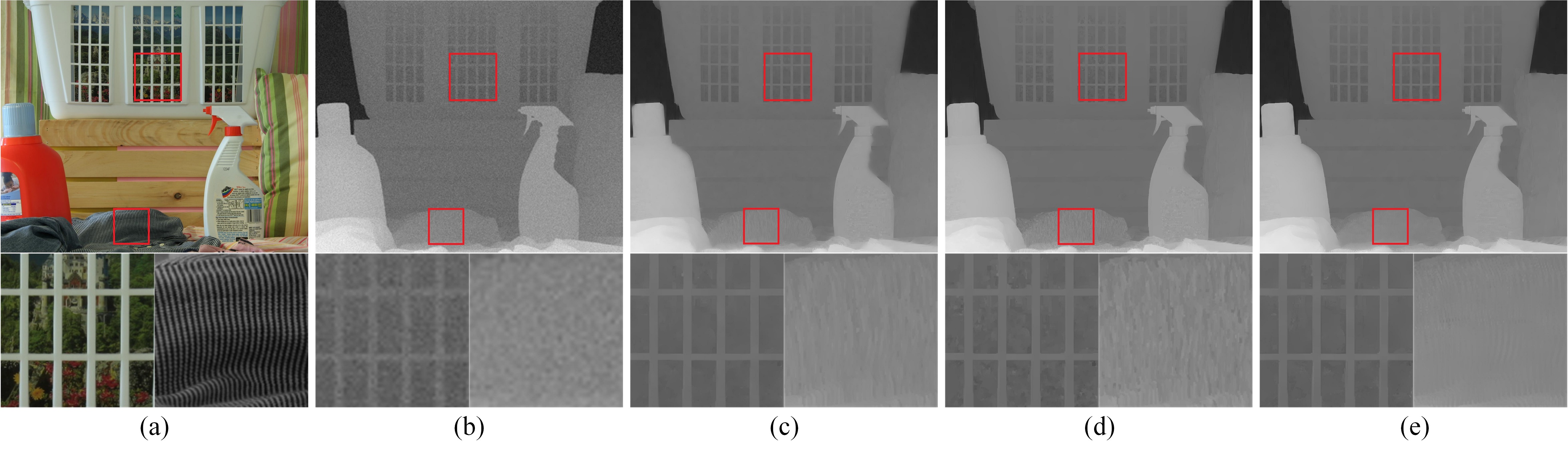}\\
  \caption{Guided depth upsampling results comparison. (a) Guidance color image. (b) Input noisy depth map (shown in bicubic interpolation). Results of (c) WLS \cite{farbman2008edge}, (d) FGS \cite{min2014fast} and (e) our SG-WLS. }\label{FigGuidedDepthUpsampling}
\end{figure*}

\noindent \textbf{Guided depth upsampling} aims at enlarging the resolution and smoothing the noise of a small noisy input depth map with the guidance of a color image. The input depth map is firstly projected onto the high resolution coordinate. We adopt Eq.~(\ref{EqSparseInterpolation}) for this task where our SG-WLS is applied to both the sparse input and the index map. The guidance weight is based on the guidance image. Results of FGS \cite{min2014fast} are obtained in a similar manner. The parameter setting of our SG-WLS is as follows: $r=4, \tau=4, \sigma_s=4, \sigma_r=3$ and $\lambda=100/200/400$ for $2\times/4\times/8\times$ upsampling. We adopt the parameters used in \cite{li2016fast} for FGS \cite{min2014fast} to produce the results. Upsampling results of different methods are shown in Fig.~\ref{FigGuidedDepthUpsampling}. The main challenges of guided depth upsampling are texture copy artifacts and blurring edges \cite{liu2017robust}. As illustrated in highlighted regions, our SG-WLS shows better performance in handing the challenges than compared methods. We further show MAD of different methods in Table~\ref{TabMAD}. Note that the results of our SG-WLS with $r=4, \tau=4$ clearly outperform the results of WLS \cite{farbman2008edge}, FGS \cite{min2014fast} and our SG-WLS with $r=1, \tau=1$. This also validates the effectiveness of using large neighborhood systems.

\begin{figure*}[h!]
\centering
  \includegraphics[width=1\linewidth]{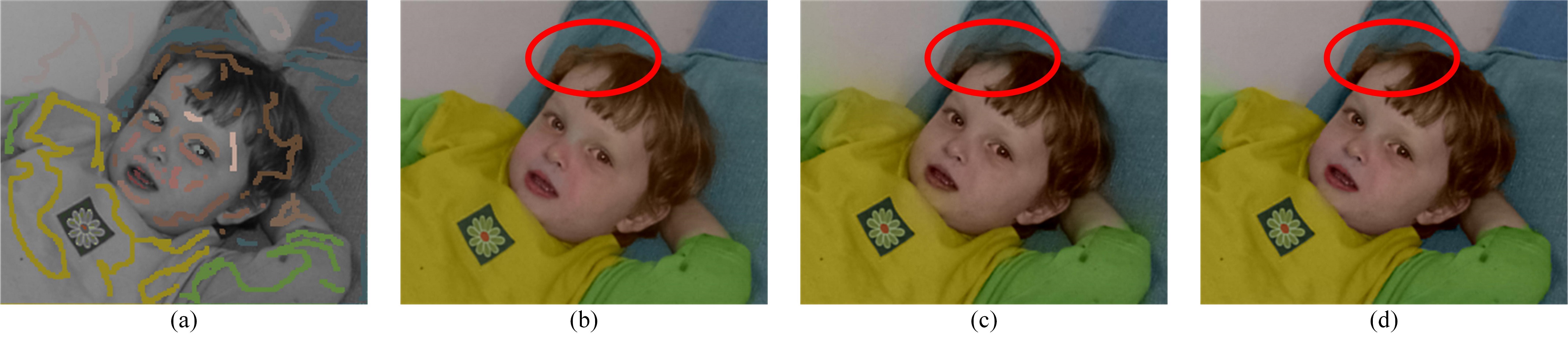}\\
  \caption{Image colorization results comparison. (a) Input gray image with color scribbles. Results of (b) Levin et al. \cite{levin2004colorization}, (c) FGS \cite{min2014fast} and (d) our SG-WLS.}\label{FigColorization}
\end{figure*}

\noindent\textbf{Image colorization} is colorizing a gray image given user specified scribbles. Two chrominance channels $U$ and $V$ extracted from the input color scribbles are used as sparse input which is propagated with the guidance of the gray image. Similar to guided depth upsampling, we use Eq.~(\ref{EqSparseInterpolation}) for the propagation of the $U$ and $V$ channel. The parameters of our SG-WLS are set as follows: $r=4, \tau=2, \sigma_s=4, \sigma_r=2, \lambda=900$. We show experimental results of different methods in Fig.~\ref{FigColorization}. Note that the hair in the red circle in the result of FGS \cite{min2014fast} is seldom colorized while the ones in our result and the result of WLS \cite{farbman2008edge} are properly colorized.

\textbf{Conclusion}
In this paper, we have presented a fast alternative approximation to the Weighted Least Squares (WLS) model,
termed Semi-Global Weighted Least Squares (SG-WLS). Both the time cost and the memory cost of our SG-WLS are much more efficient
 than that of the WLS model while it can achieve close performance to the WLS model in several applications. Our SG-WLS can overcome several limitations of previous related work due to the newly designed neighborhood system construction. Thanks to the proposed fast solution to 1D filters, our SG-WLS is also capable of a more general and larger neighborhood system other than the 4-connected/8-connected neighborhood system adopted by previous work. We show such a generalization can achieve better performance in some applications such as guided depth upsampling. Through experiments of several applications, we show the effectiveness and efficiency of our SG-WLS.

\begin{table*}
\centering
\caption{Mean Absolute Difference (MAD) of guided depth upsampling errors for different methods. Best results are in bold.}\label{TabMAD}
\resizebox{1\linewidth}{!}
{
\begin{tabular}{c|ccc|ccc|ccc|ccc|ccc|ccc}
\Xhline{1.2pt}
\multicolumn{1}{c}{\multirow{2}{*}{}} & \multicolumn{3}{|c|}{\emph{Art}} & \multicolumn{3}{c|}{\emph{Book}} & \multicolumn{3}{c|}{\emph{Dolls}} & \multicolumn{3}{c|}{\emph{Laundry} } & \multicolumn{3}{c|}{\emph{Moebius}} & \multicolumn{3}{c}{\emph{Reindeer}}\\
  \cline{2-19}
  & $2\times$ & $4\times$ & $8\times$ & $2\times$ & $4\times$ & $8\times$ & $2\times$ & $4\times$ & $8\times$ & $2\times$ & $4\times$ & $8\times$ & $2\times$ & $4\times$ & $8\times$ & $2\times$ & $4\times$ & $8\times$\\
  \Xhline{1.2pt}

  JBU \cite{kopf2007joint} & 1.59 & 2.06 & 3.18 & 0.87 & 1.24 & 2.04 & 0.91 & 1.2 & 1.98 & 0.94 & 1.38 & 2.15 & 0.89 & 1.28 & 2.05 & 0.95 & 1.36 & 2.24\\

  GF \cite{he2013guided} & 1.91 & 2.23 & 3.11 & 0.84 & 1.19 & 1.86 & \textbf{0.87} & 1.17 & 1.89 & 1.01 & 1.31 & 2.25 & 0.92 & 1.19 & 1.88 & 1.06 & 1.34 & \textbf{1.98}\\

  WLS \cite{farbman2008edge} & 1.58	& 2.52 & 3.96 & 0.9 & 1.25 & 1.85 & 0.93 & 1.3 & 1.84 & 1.03 & 1.5 & 2.33 & 0.94 & 1.34 & 1.97 &  1.09 & 1.6 &	2.42\\

  FGS \cite{min2014fast} & 1.36 & 2.01 & 3.71 & 1.25 & 1.73 & 2.58 & 1.33 & 1.89 & 2.62 & 1.11 & 1.62 & 2.61 & 1.35 & 1.95 & 2.81 & 1.46 & 2.08 & 3.07\\

  Ours ($r=1, \tau=1$) & 1.6 & 2.4 & 3.75 & 0.95 & 1.31 & 1.97 & 0.96 & 1.37 & 2.05 & 1.05 & 1.6 & 2.45 & 0.94 & 1.37 & 2.02 & 1.12 & 1.68 & 2.49\\

  Ours ($r=4, \tau=4$) & \textbf{1.26} & \textbf{1.9} & \textbf{3.07} & \textbf{0.82} & \textbf{1.12} & \textbf{1.73} & \textbf{0.87} & \textbf{1.11} &	 \textbf{1.81} & \textbf{0.86} & \textbf{1.17} & \textbf{2} & \textbf{0.82} & \textbf{1.08} & \textbf{1.79} & \textbf{0.9} & \textbf{1.32} & 2.01\\
  \Xhline{1.2pt}
\end{tabular}
}
\end{table*}

\textbf{Acknowledgements}
Correspondence should be addressed to C. Shen or J. Yang.
This work was in part supported by NSFC, China ($\#$61572315, 61503250, 61471230) and 973 Plan, China ($\#$2015CB856004).

{\small
\bibliographystyle{ieee}
\bibliography{egbib}
}

\end{document}